\documentclass[useAMS,usenatbib]{mn2e}
\pdfoutput=1
\usepackage{dsfont}
\usepackage{amsmath}
\usepackage{amstext}
\usepackage{amssymb}
\usepackage{graphicx}
\usepackage{verbatim}
\usepackage{natbib}
\usepackage{eucal}
\usepackage{calligra}
\usepackage{color}
\usepackage[overload]{textcase}
\usepackage[usenames,dvipsnames]{xcolor}
\usepackage[colorlinks,linkcolor=blue,citecolor=blue,urlcolor=blue,pdftitle={},pdfauthor={MacCrann et al.}]{hyperref}
\newcommand{\mnras}{MNRAS}
\newcommand{\jcap}{JCAP}
\newcommand{\apj}{ApJ}
\newcommand{\aj}{AJ}
\newcommand{\aap}{A\&A}
\newcommand{\prd}{Phys. Rev. D}
\newcommand{\nat}{Nature (London)}
\newcommand{\physrep}{PhR}

\newcommand{\be}{\begin{equation}}
\newcommand{\ee}{\end{equation}}
\newcommand{\bes}{\begin{equation*}}
\newcommand{\ees}{\end{equation*}}
\newcommand{\bea}{\begin{eqnarray}}
\newcommand{\eea}{\end{eqnarray}}
\newcommand{\beas}{\begin{eqnarray*}}
\newcommand{\eeas}{\end{eqnarray*}}
\newcommand{\mnuLCDM}{$m_{\nu}\Lambda$CDM}
\newcommand{\Om}{\Omega_{\rm{m}}}
\newcommand{\sigeq}{\sigma_{\rm{eq}}}

\begin{document}

\def\aj{AJ}%
\def\apj{ApJ}%
\def\apjl{ApJ}%
\def\apjs{ApJS}%
\def\apss{Ap\&SS}%
\def\aap{A\&A}%
\def\aapr{A\&A~Rev.}%
\def\aaps{A\&AS}%
\def\jcap{J. Cosmology Astropart. Phys.}%
\def\jrasc{JRASC}%
\def\mnras{MNRAS}%
\def\memras{MmRAS}%
\def\pra{Phys.~Rev.~A}%
\def\prb{Phys.~Rev.~B}%
\def\prc{Phys.~Rev.~C}%
\def\prd{Phys.~Rev.~D}%
\def\pre{Phys.~Rev.~E}%
\def\prl{Phys.~Rev.~Lett.}%
\def\pasp{PASP}%
\def\pasj{PASJ}%
\def\nat{Nature}%
\def\aplett{Astrophys.~Lett.}%
\def\physrep{Phys.~Rep.}%
\def\procspie{Proc.~SPIE}%
\let\astap=\aap
\let\apjlett=\apjl
\let\apjsupp=\apjs
\let\applopt=\ao

\defcitealias{heymans13}{HE13}
\defcitealias{takahashi12}{TA12}
\defcitealias{smith03}{SM03}

\def \seqBaseNLT{64\%~}
\def \seqBaseNLTp{(64\%)~}
\def \seqBase{54\%~}
\def \seqMnu{50\%~}
\def \seqMs{31\%~}
\def \seqIA{52\%~}
\def \seqAGN{52\%~}
\def \seqRNrun{63\%~}

\title{Cosmic Discordance: Are Planck CMB and CFHTLenS weak lensing measurements out of tune?}
\author[MacCrann et al.]{Niall MacCrann$^1$\thanks{niall.maccrann@gmail.com}, Joe Zuntz$^1$, Sarah Bridle$^1$, Bhuvnesh Jain$^{2}$, Matthew R. Becker$^{3,4}$
\vspace{0.4cm}\\
\parbox{\textwidth}{
$^{1}$Jodrell Bank Center for Astrophysics, School of Physics and Astronomy, University of Manchester, Oxford Road, Manchester, M13 9PL, UK\\
$^{2}$Department of Physics and Astronomy, Center for Particle Cosmology, University of Pennsylvania, 209 South 33rd Street, Philadelphia, PA 19104, USA\\
$^{3}$Kavli Institute for Particle Astrophysics and Cosmology, Physics Department, Stanford University, Stanford, CA 94305\\
$^{4}$Kavli Institute for Particle Astrophysics and Cosmology, SLAC National Accelerator Laboratory, Menlo Park, CA 94025\\
}}

\maketitle

\begin{abstract}
We examine the level of agreement between low redshift weak lensing data and the CMB using measurements from the CFHTLenS and Planck+WMAP polarization. We perform an independent analysis of the CFHTLenS six bin tomography results of Heymans et al. (2013). We extend their systematics treatment and find the cosmological constraints to be relatively robust to the choice of non-linear modeling, extension to the intrinsic alignment model and inclusion of baryons. 
We find that when marginalised in the $\Omega_m$-$\sigma_8$ plane, the 95\% confidence contours of CFHTLenS and Planck+WP only just touch, but the discrepancy is less significant in the full 6-dimensional parameter space of $\Lambda$CDM.
Allowing a massive active neutrino or tensor modes does not significantly resolve the tension in the full n-dimensional parameter space. Our results differ from some in the literature because we use the full tomographic information in the weak lensing data and marginalize over systematics. 
We note that adding a sterile neutrino to $\Lambda$CDM brings the 2d marginalised contours into greater overlap, mainly due to the extra effective number of neutrino species, which we find to be 0.88 $\pm$ 0.43 (68\%) greater than standard on combining the datasets. We discuss why this is not a completely satisfactory resolution, leaving open the possibility of other new physics or observational systematics as contributing factors. We provide updated cosmology fitting functions for the CFHTLenS constraints and discuss the differences from ones used in the literature.
\end{abstract}


\section{Introduction}

The Cosmic Microwave Background (CMB) radiation has been the most powerful probe of cosmology for more than a decade. The Planck satellite \citep{planck13} gives us an unprecedented view of the temperature fluctuations at recombination and 
the Wilkinson Microwave Anisotropy Probe \citep[WMAP, ][]{bennett97} has until recently \citep{planckcosmo15} provided the most detailed maps of the polarisation fluctuations \citep{bennett13}. 
Planck and WMAP polarisation together provide a self-consistent constraint on the 6 parameter $\Lambda$CDM cosmological model i.e. a flat universe containing only cold dark matter and baryons, and a cosmological constant, $\Lambda$.

At the same time, pressure is mounting on the $\Lambda$CDM model from tension between the CMB and low-redshift measurements of matter clumpiness. The primary CMB anisotropies place a constraint on the matter fluctuation amplitude at the time of recombination, which can be extrapolated to the present day for a particular assumed cosmological model. The primary measures of the amplitude of matter fluctuations at low redshift are weak lensing,
galaxy clustering and the abundance of galaxy clusters. Low-redshift observations seem to be finding a lower value for this fluctuation amplitude than expected in $\Lambda$CDM \citep{beutler14b, plancksz13, vikhlinin09}.  This could be reconciled by new physics which reduces the rate of clustering between recombination and today
\citep{planck_sz,hamann13,battye14,beutler14b,dvorkin14,archidiacono14}.

Gravitational lensing is the most direct method for measuring the distribution of matter in the 
low-redshift 
universe. The image distortion of distant galaxies in typical patches of sky was first detected in 2000 \citep{bacon00, kaiser00, vanwaerbeke00, wittman00} and last year the Canada-France-Hawaii Telescope Lensing Survey (CFHTLenS) provided the tightest constraints on cosmology yet from cosmic shear. This arguably provides one of the most robust and constraining low-redshift measures of cosmology, and thus this is the low-redshift dataset we focus on in this paper. 

One way to reduce the matter clustering rate is for some of the matter to travel fast enough to leave the clumps and smear out the fluctuations (``free-streaming''). Active neutrinos are an obvious candidate for this hot dark matter, because we already know they have mass \citep{beringer12} and particle physics experiments allow a mass range that would have a significant impact on cosmology \citep{lobashev99,weinheimer99}. They have been invoked at various times to reconcile CMB and low-redshift counts of galaxy clusters.

Even if the active neutrino has the smallest mass allowed by particle physics experiments, an alternative hot dark matter particle might be responsible for smearing out the fluctuations. A sterile neutrino is a promising candidate which would also affect the CMB anisotropies by introducing an additional relativistic species in the early universe. 

In this paper we focus in detail on combining the CMB with the CFHTLenS low-redshift dataset to examine whether they alone warrant new physics. In contrast to the earlier papers, we use the full 6 tomographic redshift bins and marginalise over intrinsic alignments, as in \cite{heymans13} and described in Section 2. Earlier papers drew conclusions about agreement between datasets by comparing marginalised contours in one or two dimensions. In Section 3 we investigate whether these conclusions hold up in the full multi-dimensional parameter space, and extend the treatment of weak lensing systematics.
In Section 4 we investigate the effect of cosmological extensions (massive active neutrinos, a massive sterile neutrino, tensors and running of the spectral index) to the base model. 
We compare with related work, and discuss other possible explanations for the tension in Section 5. 

\section{Datasets and methodology}

The Planck satellite \citep{planck13} provided high resolution ($\sim$10 arcminutes) temperature maps of the CMB at a range of frequencies between $\sim$25 and $\sim$1000 GHz. These observations allow the estimation of the CMB temperature power spectrum for $2\le l \le 2500$ \citep{planck_like}. We use the publicly available Planck likelihood codes which use this power spectrum, and the corresponding polarisation power spectrum from WMAP9 \citep{bennett13}. Throughout, we marginalise over the 14 nuisance parameters which account for astrophysical systematics in the Planck likelihood codes. We refer to this combination as Planck+WP. 

The Canada France Hawaii Lensing Survey ~\citep{heymans12}, hereafter referred to as CFHTLenS, is a 154 square degree multi-filter survey which achieved an effective weighted number density of 11 galaxies per square arcminute with shape and photometric redshift estimates. \cite{kilbinger13} performed a 2d cosmic shear analysis of the CFHTLenS data, producing constraints on the $\Lambda$CDM model which they approximated by $\sigma_8 (\Omega_{\rm m}/0.27)^{0.59}=0.787\pm0.032$. 
\cite{heymans13} (\citetalias{heymans13} henceforth) performed a tomographic cosmic shear analysis of the CFHTLenS data, dividing the galaxies into six tomographic redshift bins (with photometric redshift estimate between 0.2 and 1.3) and taking into account the effect of galaxy intrinsic alignments using a free parameter for the overall intrinsic alignment amplitude. For each tomographic bin combination, they measured the real space shear-shear correlation functions $\xi_{+,-}(\theta)$ in 5 evenly log-spaced angular bins for $1\le \theta \le 40$ arcmin. In this paper we use the full \citetalias{heymans13} correlation functions and covariance matrices (which were obtained from N-body simulated mock surveys), and marginalise over the same model for intrinsic alignments as in \citetalias{heymans13}. 

The analysis in this paper is performed with CosmoSIS, a new cosmological parameter estimation framework (Zuntz et al. 2014 in prep).  A parameter estimation problem in CosmoSIS is represented as a sequence of independent modules each performing a specific part of the calculation and passing on their results to later modules.  For this work the modules were: {\sc camb} \citep{lewis99}, to calculate CMB and linear matter power spectra and expansion histories; {\sc Halofit} for non-linear power; a module based on {\sc cosmocalc}\footnote{https://bitbucket.org/beckermr/cosmocalc} to compute cosmic shear spectra and intrinsic alignments; a custom module to compute the 2-point shear correlation functions $\xi_\pm(\theta)$ from $C_\ell$; the commander, lowlike and CAMSpec Planck likelihood codes \citep{planck_like}; and a custom CFHTLenS likelihood code. 
As a default we use the {\sc Halofit} formulation as implemented by {\sc camb}, which is \cite{takahashi12} with modified massive neutrino parameters (although we compare this nonlinear correction to others in Section \ref{sec:nl}). We'll refer to this implementation as \citetalias{takahashi12} from now on. Note that \citetalias{heymans13} used the fitting formula of Eisenstein \& Hu to get the linear matter power spectrum and the \cite{smith03} (SM03 henceforth) version of {\sc Halofit} to perform the non-linear correction.

We use the following parameter definitions. 
$\Om$ is the total matter density at redshift zero (as a fraction of the critical density at redshift zero).
The present-day baryon density is given by $\Omega_b$.
$\sigma_8$ is the rms flucutation in 8$h^{-1}\: \rm{Mpc}$ spheres at the present day in linear theory.
The spectral index of the scalar primordial power spectrum is given by $n_s$.
$\tau$ is the optical depth 
due to 
reionization.
The Hubble constant is written as $h$, in units of $100\: \rm{(km/s)/Mpc}$. 
When we refer to `base $\Lambda$CDM', we mean the same model as the \cite{planck_cosmopar} baseline model - the normal 6 parameter $\Lambda$CDM model, assuming 1 massive active neutrino eignestate, with $m_{\nu}=0.06\: \rm{eV}$. 

\section{Discordance in $\Lambda$CDM}\label{sec:data}

\begin{figure*}
\includegraphics[width=8.8cm]{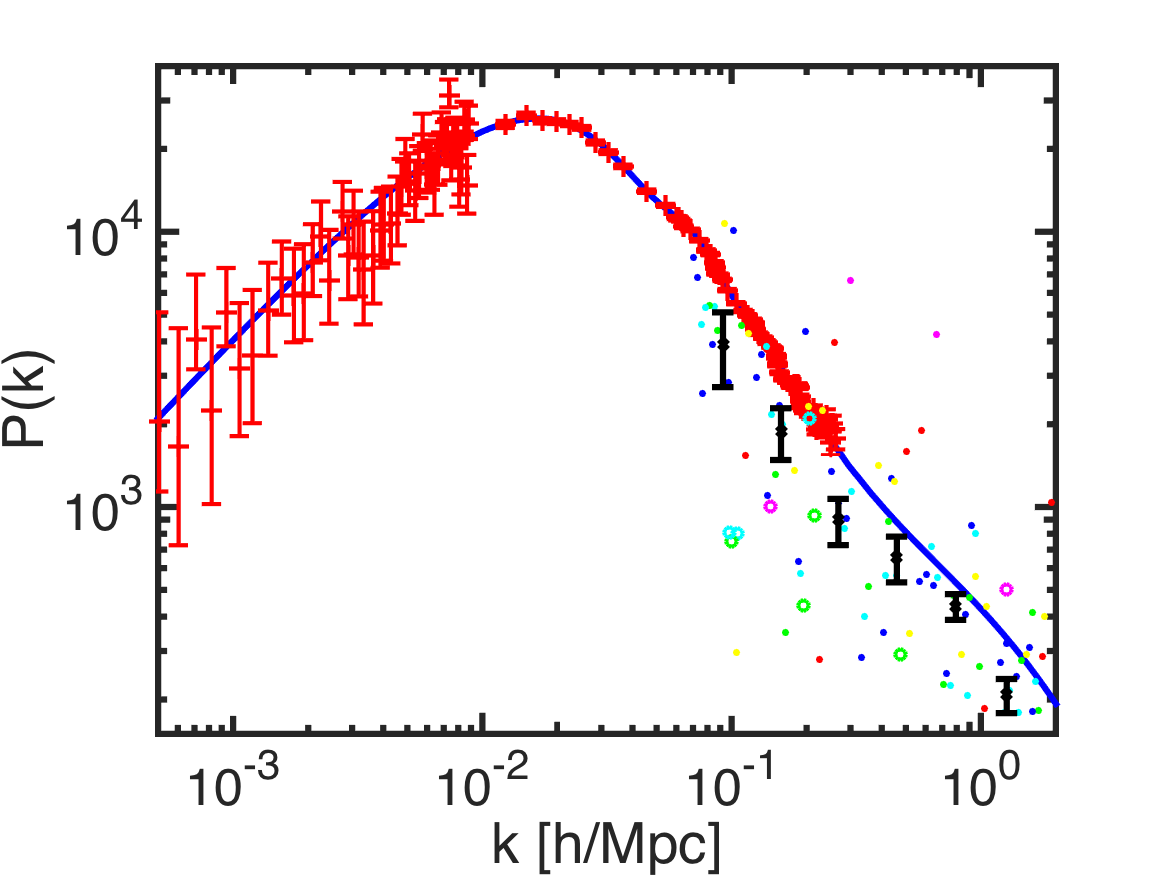}
\includegraphics[width=8.8cm]{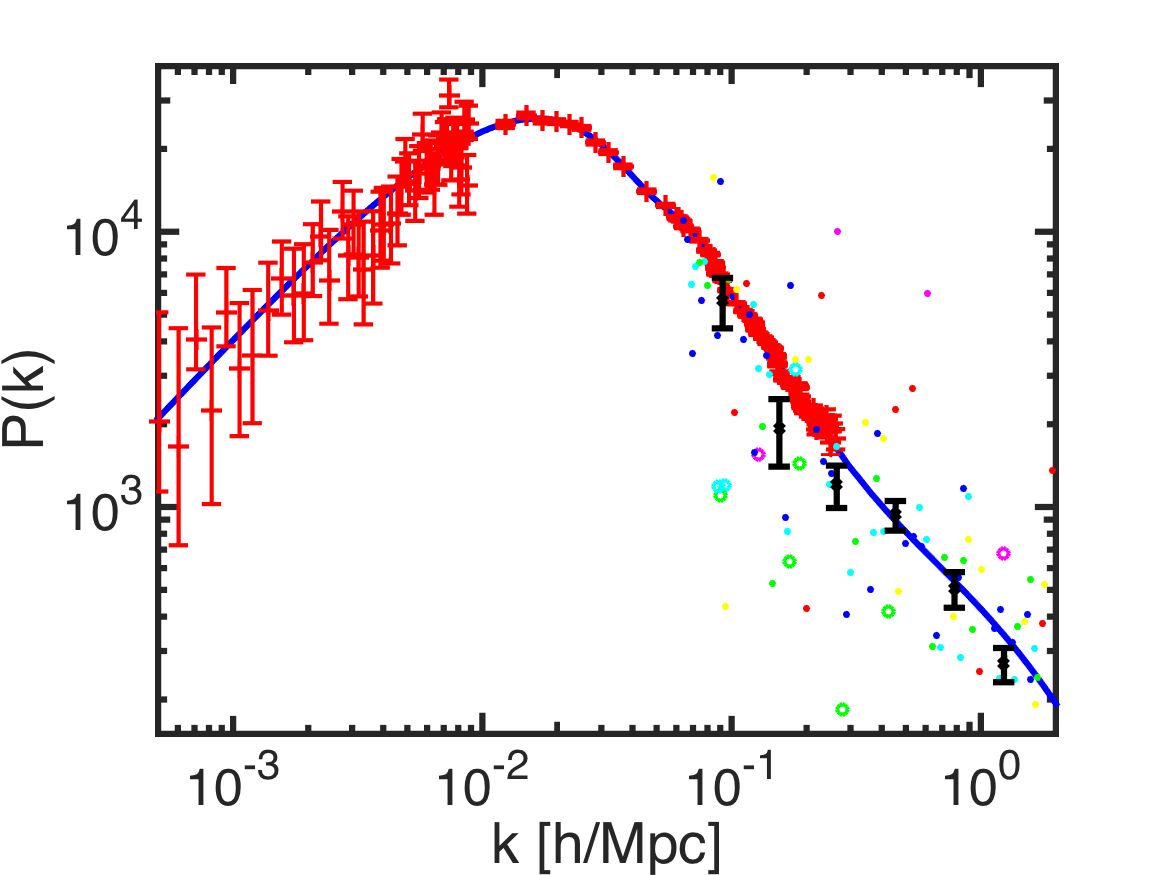}
\caption{The Planck and CFHTLenS data superposed onto the present day matter power spectrum, using the method of \protect\cite{tegmark02}. Each coloured CFHTLenS point corresponds to an angular correlation function measurement. Cross correlations with tomographic bin 1 are magenta, with bin 2 (and not with bin 1) are red, with bin 3 (and not with bins 1 or 2) are yellow, bin 4 are green, bin 5 are cyan and bin 6 are blue. There are 105 points from CFHTLenS $\xi_{+}$ which are illustrated by the coloured points. Some of them fall at smaller scales than shown on this plot, and some are negative and shown by an open circle.
These have been averaged using the noise covariance matrix to make the black points. 
Left: For the Planck best fit cosmology. Right: For the Planck + CFHTLenS $\Lambda$CDM best fit cosmology. Note that because of the extrapolation to the matter power spectrum, both the points and the lines move when the cosmology changes. In the range of the CFHTLenS data points the line moves down by about the same amount as the CFHTLenS points move up, on switching the cosmology from Planck (left panel) to Planck + CFHTLenS (right panel).}
\label{fig:cmb_cl}
\end{figure*}

We assess the level of agreement between CFHTLenS and Planck+WP in the 6 parameter 
base 
$\Lambda$CDM model, and provide an updated fitting function to the CFHTLenS data.

\subsection{Quantifying the tension}\label{sec:tension}

\begin{figure}
\includegraphics[width=\linewidth]{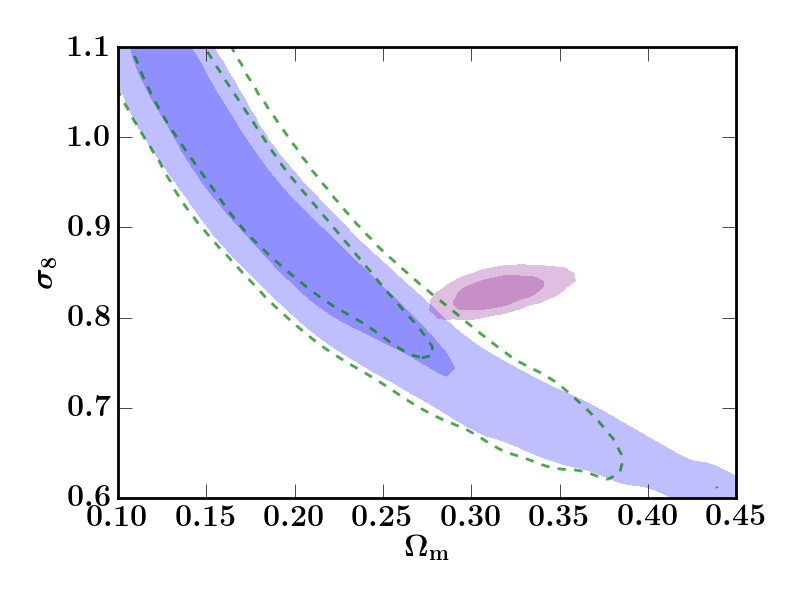}
\caption{Constraints in the clustering amplitude $\sigma_8$ and dark matter density $\Om$ plane from Planck+WP and CFHTLenS, 
assuming our base cosmological model. 
Filled blue banana: 1 and 2$\sigma$ CFHTLenS only constraints using all $\theta$ bins in $\xi_{+/-}(\theta)$. Dashed green banana: 1 and 2$\sigma$ CFHTLenS only constraints excluding small scales (see Section \ref{sec:nl} for cuts on $\theta$).
Small, purple contours: The constraints on the base model from Planck+WP. Discrepancy between the 2d marginalised Planck+WP and CFHTLenS contours is clear at 
the $\sim95\%$ level}
\label{fig:om_s8_base}
\end{figure}

Fig.~\ref{fig:cmb_cl} shows the Planck and CFHTLenS data superposed onto the matter power spectrum \citep[following][]{tegmark02}.
This can provide a qualitative indication of the level of agreement between the datasets and a given cosmological model. However, note that the conversion of observables to the matter power spectrum is highly dependent on the assumed cosmology 
. Therefore an apparent disagreement between two datasets can simply be an indication that the wrong model was assumed when converting the data points. 
In the left panel, we can see that the Planck best-fit cosmology goes straight through the Planck datapoints, as expected from the good fit of the Planck Cls to the best fit theory model. The CFHTLenS datapoints appear to be more often below the theory line, as expected from the lower preferred $\sigma_8$. This was also illustrated in \cite{battye14} using the cosmic shear correlation function.

In Fig.~\ref{fig:om_s8_base} we show that the two-dimensional marginalised constraints from Planck+WP and CFHTLenS are discrepant in the $\Om$-$\sigma_8$ plane: the 2$\sigma$ contours 
only just
touch. This is a significantly stronger conclusion than reached in other works, e.g. \cite{leistedt14}, \cite{beutler14b}.
There are four main reasons for this: (i) We use an improved non-linear {\sc Halofit}  treatment (see Section \ref{sec:ff} below) for the lensing constraints.
(ii) we use exactly the same cosmological model (i.e. include an active neutrino with mass 0.06 eV) for the CFHTLenS constraints as for the Planck+WP constraints, although Figure 4 of \cite{beutler14b} suggests that at least for the CFHTLenS constraints, this is less important than (i).
(iii) we 
use a full likelihood analysis rather than just a prior in the $\Om$-$\sigma_8$ plane;
(iv) we follow \citetalias{heymans13} by using 6 bin tomographic results marginalised over intrinsic alignments (see Fig. 4 of \citetalias{heymans13} for the effect of this). 

However, the fact that the 2d marginalised CFHTLenS and Planck+WP contours do not overlap 
is not necessary or sufficient to prove 
that they are discrepant, since the $\Lambda$CDM model has 6 dimensions, so it is the amount of overlap in 6 dimensions that is important. 

One way we can quantify the 
discrepancy between two 
datasets (e.g. Planck+WP and CFHTLenS) 
within a particular $n$-parameter cosmological model is by 
checking how much the $n$-dimensional posterior distributions overlap. 
We first calculate the positions of the 68\% and 95\% surfaces of equal likelihood in the full n-dimensional parameter space
for a given dataset and can then assess whether a given point 
lies within these confidence intervals. We call these surfaces iso-likelihood surfaces, or `iso-likes' for brevity.
More generally we can identify the percentage iso-like a given point in parameter space lies on, for each dataset. 

We find the multi-dimensional iso-likes as follows.
We perform fits of the model to the two measurements individually, allowing us to obtain a histogram of 
probability 
values for each dataset. 
As in 
the 1d case, we define the $68\%$ iso-like as the surface of equal likelihood which contains 68\% of the probability 
distribution, or in the case of MCMC samples, 68\% of the samples. 
This allows us to identify the 
probability
value of the 68\% iso-like for each dataset. 
More generally, we can use this histogram to read off the percentage iso-like for any point in parameter space, given its 
probability 
value.

As an example of a point of interest, we 
perform a 
joint fit, and define  $\sigma_i(\mathbf{p}_{\textrm{joint}})$ as the percentage iso-like on which the joint-best fit point lies, for dataset $i$ (where $i$ is one of C and P, denoting CFHTLenS and Planck+WP respectively). 
The values of $\sigma_{\rm{C}}(\mathbf{p}_{\textrm{joint}})$ and $\sigma_{\rm{P}}(\mathbf{p}_{\textrm{joint}})$ are given in Table 1.
The best joint fit is a poor fit to CFHTLenS, lying on the 76\% iso-like. The fact that it is an acceptable fit to Planck+WP reflects the greater constraining power from Planck+WP, 
which pulls the best fit point close to the best fit to Planck+WP alone.

We also wish to know if there are regions of parameter space which are a good fit to both datasets (albeit a slightly worse fit to Planck+WP), i.e. the minimum percentage iso-likes which overlap. For this we define $\sigeq$, the minimum value of $\sigma_{\rm{C}}$=$\sigma_{\rm{P}}$. 
Therefore 
$\sigeq$ is the best percentile value for which equal percentage iso-likes of Planck+WP and CFHTLenS touch. For base $\Lambda$CDM we find $\sigeq=\seqBase$, (or $\sigeq=\seqBaseNLT$ when cutting small scales from the CFHTLenS correlation functions, see Section \ref{sec:nl}). This means that the best points, or at least those where the tension is least, are still on at least the \seqBase (\seqBaseNLT) iso-likes of \textit{both} probes. These $\sigma$ values are collected for this and subsequent sections in Table \ref{tab:res}.

It's clear that when considering the relative positions of the 68\% and 95\% confidence intervals as a test of tension or discrepancy, then the number of dimensions under consideration is important - a naive interpretation is that the marginalised 2d picture suggests a greater tension than the 6d case. However this is likely to be largely a geometrical effect - when we marginalise over some parameters, if those parameters are (even weakly) constrained, the surface of equal probability containing e.g. 68\% of the probability (the 68\% contour in 2d) will be found at a higher probability, making confidence regions tighter. Appendix~\ref{appendix:nd_contour_stuff} illustrates this effect further for the case of two gaussian probability distributions. We believe that while the 2d marginalised picture does still give a useful indication of the tension, the full n-dimensional $\sigma_{eq}$ should also be considered as as an alternative and more conservative assessment.

\begin{table*}

\begin{tabular}{ r | c | c | c | c | c | l } 
Model &  Datasets  & small scales cut? & $\sigma_C(\mathbf{p}_{\textrm{joint}})$ & $\sigma_P(\mathbf{p}_{\textrm{joint}})$ & $\sigeq$ \\
\hline
\hline
$\Lambda$CDM (6) & P + C & No & 76\% & 23\% & \seqBase & \\
\hline
$\Lambda$CDM (6) & P + C & Yes & 89\% & 11\% & \seqBaseNLT & \\
\hline 
$\Lambda$CDM + IA(z) (6+1) & P + C & Yes & 70\% & 15\% & \seqIA & \\
\hline 
$\Lambda$CDM + AGN (6+1) & P + C & Yes & 86\% & 8\% & \seqAGN & \\
\hline 
$m_{\nu}\Lambda$CDM (6+1) & P +  C & Yes & 90\% & 1\% & \seqMnu\\
\hline
$m_{\rm{s}}^{\rm{eff}}\Delta N_{\rm eff}\Lambda$CDM (6+2) & P +  C & Yes & 60\% & 2\% & \seqMs\\
\hline
$r\alpha_{run}\Lambda$CDM (6+2) & P + C & Yes & 92\% & 8\% & \seqRNrun\\
\hline

\end{tabular}
\caption{Goodness of fit of joint fit to individual datasets, for several extensions to $\Lambda$CDM. `C' and `P' denote CFHTLenS and Planck+WP respectively. `small scales cut' refers to removing some of the $\xi_{+/-}(\theta)$ bins used in the CFHTLenS analysis, as described in Section \ref{sec:nl}. $\mathbf{p}_{\textrm{joint}}$ denotes the parameters of the best joint fit to the datasets. $\sigma_i$ values are defined at the end of Section \ref{sec:tension}. In parentheses after the model names, we also include the number of parameters in the model, not including the Planck nuisance parameters.}
\label{tab:res}
\end{table*}

\subsection{Sensitivity to the choice of nonlinear matter power spectrum}\label{sec:nl}

\begin{figure}
\includegraphics[width=\linewidth]{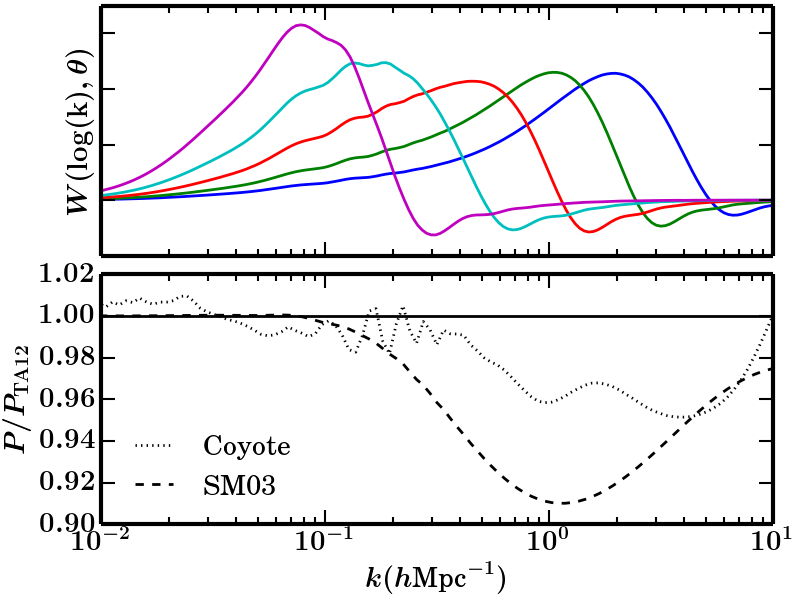}
\caption{Top panel - the weight functions, $W(log_{10}(k),\theta)$ for $\xi_+(\theta)$, for the autocorrelation of the highest redshift CFHTLenS bin. The weight functions give the relative contribution to $\xi_+(\theta)$ as a function of $k$. The 5 lines are for the 5 $\theta$ bins used (with bin centres at 1.65, 3.58, 7.76, 16.80, 36.18 arcmin), with lower $\theta$ bins peaking at higher $k$. Bottom panel - the nonlinear matter power spectrum at z=0.5 predicted by Coyote and \citetalias{smith03}, as a fraction of the \citetalias{takahashi12} prediction.}
\label{fig:Pk}
\end{figure}

The strength of weak lensing lies in its ability to constrain the matter power spectrum, $P_{\delta}(k)$ however, it is most sensitive to scales where nonlinear effects on $P_{\delta}(k)$ are significant. This is demonstrated by Fig.~\ref{fig:Pk}, the upper panel of which shows the weighting of $k$ scales in $\xi_+$ for the autocorrelation of the highest redshift CFHTLenS redshift bin. We show $W(\rm{log}(k),\theta)$, where
\begin{equation}
\xi_+(\theta) = \int \rm{dlog}(k) W(log(k),\theta) P_{\delta}(k)
\end{equation}
The 5 lines are the 5 angular bins ($1\le\theta\le40$ arcmin) used in the \citetalias{heymans13} measurement, with larger angles peaking at lower $k$.
The lower panel shows the fractional difference between the two {\sc Halofit} versions, \citetalias{smith03} and \citetalias{takahashi12} and the prediction of the publicly available code FrankenEmu\footnote{http://www.hep.anl.gov/cosmology/CosmicEmu/emu.html}, a matter power spectrum emulator based on the Coyote Universe simulations \citep{heitmann14}, which we'll refer to as Coyote.
Comparing the two panels, it is clear, particularly for smaller angular bins, that the choice of nonlinear correction is important for the $k$-scales being probed.


\cite{beutler14b} already noted a $\approx 1\sigma$ shift in the constraint on $\sigma_8$ from using the newer version of {\sc Halofit}. This is not unexpected when we look at the fractional differences in P(k) for different nonlinear prescriptions, shown in Fig.~\ref{fig:Pk}. \citetalias{heymans13} suggest the conservative approach of cutting some of the lower $\theta$ bins in $\xi_{+/-}$ as a way of reducing the importance of the nonlinear correction. They boost and decrease the \citetalias{smith03} non-linear correction by $\pm 7\%$, and propose cutting all $\theta$ bins where the predicted $\xi$ changes by more than 10\%. For $\xi_+$, this corresponds to $\theta \le 3 $ arcmin for tomographic bin combinations including bins 1 and 2. For $\xi_-$  (which is sensitive to higher $k$ than $\xi_+$ for a given angular scale), this corresponds to $\theta \le 30 $ arcmin for tomographic bin combinations including bins 1, 2, 3 and 4, and $\theta \le 16 $ arcmin for tomographic bin combinations including bins 5 and 6. 

\begin{figure}
\includegraphics[width=\linewidth]{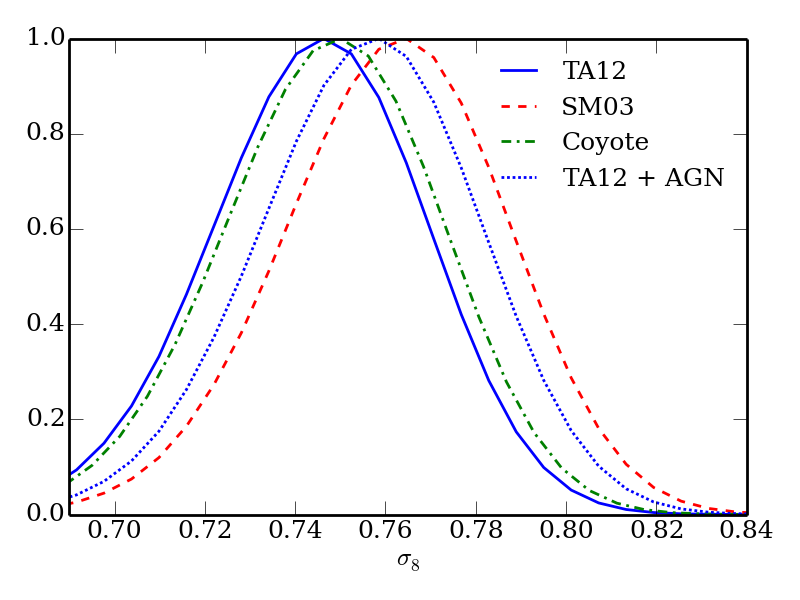}
\caption{CFHTLenS $\sigma_8$ constraints in an otherwise fixed fiducial cosmology, with 3 different nonlinear power spectrum treatments. Also shown is the constraint using \citetalias{takahashi12} and a prescription for AGN feedback described in Section \ref{sec:bar}}
\label{fig:sig8}
\end{figure}

We adopt this scheme for the rest of the paper, and perform the following simple test on the sensitivity to the choice of nonlinear correction: We fix all parameters except $\sigma_8$ to the best joint-fit Planck+WP and CFHTLenS cosmology, and obtain 1d CFHTLenS constraints on $\sigma_8$, for each of \citetalias{takahashi12}, \citetalias{smith03} and Coyote. Fig.~\ref{fig:sig8} shows the results of this test. Even after implementing the conservative $\theta$ cut, there is still a $0.7\sigma$ shift between \citetalias{smith03} and \citetalias{takahashi12}, although the constraints from \citetalias{takahashi12} and Coyote are very similar for this slice of parameter space. Note that since we've fixed all other parameters, the errorbar on $\sigma_8$ will be smaller than when marginalising over e.g. $\Omega_m$, so in some sense this is a conservative test. Encouraged by this, we continue using \citetalias{takahashi12} for the rest of the paper, since it can be used consistently with non-zero neutrino mass.

\subsection{Baryonic feedback}\label{sec:bar}

The matter power spectrum (and therefore the weak lensing convergence power spectrum) is also affected by baryonic feedback at $k>1h\:\rm{Mpc}^{-1}$, as pointed out by \cite{white04} and \cite{zhan04} who, using simple models of the effect, reported up to several percent changes in the convergence power spectrum at $l>1000$. \cite{jing06}, \cite{rudd06}, \cite{guillet10}, \cite{casarini12} all confirmed the significance of baryonic feedback at $k>1h\:\rm{Mpc}^{-1}$ by comparing hydrodynamical simulations to pure N-body dark matter ones.
\cite{shaye10} performed the OverWhelmingly Large Simulations (OWLS) to investigate the effect
of several different baryonic effects on the cosmic star formation history, while \cite{vandalen11} used OWLS to investigate the effect of baryons on the matter power spectrum. It has been argued (\cite{vandalen11}, \cite{sembolini11}) that  the OWLS `AGN' model, which accounts for the presence of black holes and AGN feedback in dark matter halos using the prescription described in \cite{booth09}, is the most realistic of the different OWLS models, since it matches well both the observed optical and X-ray properties of galaxy groups (\cite{mccarthy11}). To test the effect of AGN feedback on the CFHTLenS constraints, we use the matter power spectra\footnote{http://www.strw.leidenuniv.nl/VD11/} derived by \cite{vandalen11} from the OWLS. 
Specifically, we use the `AGN' spectrum which we call $P_{\delta}^{\rm{AGN}}$ and the `DMONLY' spectrum $P_{\delta}^{\rm{DMONLY}}$, which is the power spectrum derived from the OWLS dark matter only simulation. We approximate the effect of AGN feedback by multiplying our varying \citetalias{takahashi12} power spectrum, $P_{\delta}^{\rm{TA12}}$ by the ratio of the `AGN' power spectrum to the `DMONLY' power spectrum

\begin{equation}
P_{\delta} = \frac{P_{\delta}^{\rm{AGN}}}{P_{\delta}^{\rm{DMONLY}}} P_{\delta}^{\rm{TA12}} \, .
\end{equation}

\begin{figure*}
\includegraphics[width=\linewidth]{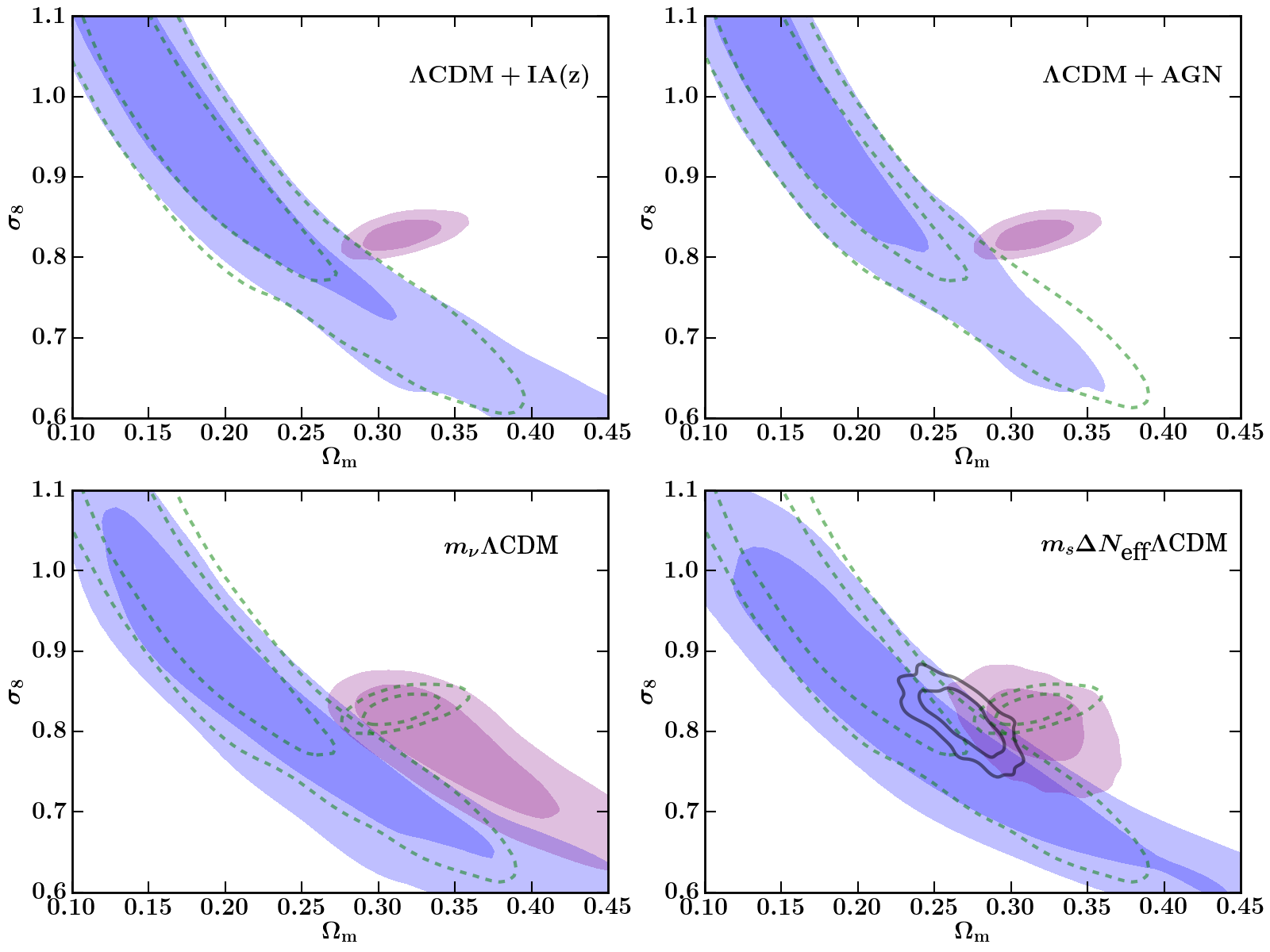}
\caption{68\% and 95\% confidence regions in the clustering amplitude $\sigma_8$ and dark matter density $\Om$ plane from Planck+WP alone and CFHTLenS alone. In all panels, dashed green contours represent the base $\Lambda$CDM constraints from Fig.~\ref{fig:om_s8_base}.
\textbf{Top left}: CFHTLenS with extra IA redshift scaling parameter (filled blue contours) and Planck+WP (smaller purple contours).
\textbf{Top right}: CFHTLenS marginalised over an AGN feedback parameter (filled blue contours) and Planck+WP (smaller purple contours).
\textbf{Bottom left}: CFHTLenS (blue contours) and Planck+WP (smaller purple contours) allowing varying active neutrino mass.
\textbf{Bottom right}: CFHTLenS (blue contours) and Planck+WP (smaller purple contours) allowing a  massive sterile neutrino. The black contours are the joint fit.} 
\label{fig:om_s8_multi}
\end{figure*}

 We obtain a 1d constraint on $\sigma_8$ as before, 
shown by 
 the dotted line labelled TA12 + AGN in Fig.~\ref{fig:sig8}. Largely due to the fact that we have cut the smallest scales from our analysis, the shift from introducing AGN feedback is smaller than the shift arising from the use of different Halofit versions, 
which suggests 
 that the effect of AGN feedback is probably not important here. Nevertheless, we also repeat the analysis of Section \ref{sec:tension}, introducing a new parameter $\alpha_{\rm{AGN}}$ given by
 \begin{equation}
P_{\delta} =\left(1 + \alpha_{\rm{AGN}} \frac{(P_{\delta}^{\rm{AGN}} - P_{\delta}^{\rm{DMONLY}})}{P_{\delta}^{\rm{DMONLY}}}\right) P_{\delta}^{\rm{TA12}} \, .
\end{equation}
Thus we allow the strength of the AGN feedback to vary by allowing $\alpha_{\rm{AGN}}$ to vary. The top right panel of Fig.~\ref{fig:om_s8_multi} shows the effect on the CFHTLenS alone contours when we allow $0< \alpha_{\rm{AGN}}<2$. 
We repeat the analysis of Section \ref{sec:tension}, but obtain only a small improvement in agreement between the two probes, with $\sigeq =$\seqAGN.
In the joint fit, we allow  $-3< \alpha_{\rm{AGN}}<3$, and find a preferred value of $\alpha=0.78^{+1.5}_{-1.02}$,. The fact there is only weak preference for positive $\alpha_{\rm{AGN}}$ is consistent with this parameter not being very helpful in resolving the tension. 

This is the first combined CMB and lensing analysis to constrain baryonic feedback and cosmology simultaneously, although we note that this is a very simplistic prescription for AGN feedback, let alone baryonic feedback as a whole. \cite{harnois-deraps14} use three of the OWLS to construct a fitting function for the effect of baryonic feedback on the power spectrum, and use CFHTLenS to constrain this model while fixing the cosmology. Unlike this analysis, they extend the CFHTLenS data to sub-arcminute scales, and find indications of a preference for a universe with baryonic feedback.  \cite{eifler14} show the importance of accounting for baryonic feedback for stage III and IV weak lensing experiments, and propose a PCA marginalisation approach that uses information from a range of hydrodynamical simulations, as a way of removing the bias with 3 or 4 nuisance parameters.


\subsection{Sensitivity to the IA model}\label{sec:IA}

We follow \citetalias{heymans13} by using the non-linear linear alignment model (henceforth NLA model, \cite{bridle07}) to account for intrinsic alignments. The NLA model is based on the linear alignment model \citep{catelankb01,hirata04}, which assumes that galaxies are aligned with their haloes which are in turn are aligned with the local tidal gravitational field; for a given redshift the intrinsic galaxy ellipticity is taken to be proportional to the linear theory tidal field strength. In the linear alignment model, the intrinsic-intrinsic (II) and shear-intrinsic (GI) power spectra are given by
\begin{equation}\label{eqn:LA}
P_{\rm{II}}(k,z) = F^2(z)P_{\delta}(k,z) ,\;\;   P_{\rm{GI}}(k,z) = F(z)P_{\delta}(k,z),
\end{equation}
\noindent
where
\begin{equation}\label{eqn:fz}
F(z) = -AC_1\rho_{\textrm{crit}}\frac{\Omega_{\rm m}}{D(z)} .
\end{equation}
\noindent
$\rho_{\rm crit}$ is the critical density at $z=0$, $C_1 = 5 \times 10^{-14}h^{-2}M_{\odot}^{-1}{\rm Mpc}^3$, and $A$, the dimensionless amplitude, is the single free parameter.

In the NLA model, the linear matter power spectrum in equation \ref{eqn:LA} is replaced with the non-linear matter power spectrum. One of the main uncertainties of both of these alignment models is the redshift scaling - it may be that alignment was produced at high redshift during galaxy formation, but the strength of the signal is likely to have evolved over cosmic time.
Hence we try a simple extension to the NLA model, by introducing a power law redshift scaling, $\alpha_{\rm{IA}}$, so that
\begin{equation}
F(z) = -AC_1(1+z)^{\alpha_{\rm{IA}}}\rho_{\textrm{crit}}\frac{\Om}{D(z)}.
\end{equation}

We repeat the analysis of Section \ref{sec:tension}, and obtain the marginalised constraints in the $\Om$-$\sigma_8$ plane shown in the top left panel of Fig.~\ref{fig:om_s8_multi}. A small shift in the CFHTLenS contours is apparent when including the extra intrinsic alignment parameter, but by eye it does not appear significant in resolving the tension with Planck+WP. 
We find $\sigeq =$\seqIA for this model, which supports this conclusion.


In agreement with \citetalias{heymans13} we find that negative values of the intrinsic alignment amplitude parameter $A$ are slightly preferred for $\alpha_{\rm{IA}}=0$. 
We allow a prior range of $-5<\alpha_{\rm{IA}}<5$ and find $\alpha_{\rm{IA}}$ to be unconstrained but preferred to be strongly negative for both CFHTLenS alone and for CFHTLenS + Planck+WP. This can be understood from the relatively large amount of power at low redshift in the CFHTLenS data - see \citetalias{heymans13} Fig. 2. The negative power law index allows more intrinsic alignment contribution at low redshift and very little at high redshift. An even more negative intrinsic alignment amplitude $A$ is preferred than for $\alpha_{\rm{IA}}=0$, for both CFHTLenS alone and CFHTLenS + Planck+WP. This makes the contribution from the dominant intrinsic alignment term (GI) positive, to match the relative excess of power in the observations at low redshift.


\subsection{A New CFHTLenS Fitting Function}\label{sec:ff}

\citetalias{heymans13} presented the constraint $\sigma_8(\Om/0.27)^{0.46} = 0.774^{+0.032}_{-0.041}$ that has been used in combination with other datasets instead of running a full likelihood analysis. The \citetalias{heymans13} analysis used the \cite{smith03} version of {\sc Halofit}.  \cite{beutler14b} showed a $\sim1\sigma$ reduction in the  \cite{kilbinger13} CFHTLenS constraint on $\sigma_8$ at $\Om=0.3$ when using a newer {\sc Halofit} version, 
however, that analysis used angular scales down to 0.9 arcmin in both $\xi_+$ and $\xi_-$, which are highly sensitive to the nonlinear modelling. From our more conservative analysis, we find
\begin{equation}
\sigma_8(\Om/0.27)^{0.48} = 0.768\pm0.037, 
\end{equation}
which is slightly lower than, but consistent with the \citetalias{heymans13} result for e.g. $\Omega_m=0.27$.

\section{Discordance in extensions to $\Lambda$CDM}

In this section we try the following extensions to $\Lambda$CDM: massive active neutrinos, a massive sterile neutrino, and primordial tensor modes, and quantify how much they resolve the tension between Planck+WP and CFHTLenS.

\subsection{Discordance in \protect\mnuLCDM}


We wish to know whether allowing a greater active neutrino mass 
alleviates 
the tension between Planck+WP and CFHTLenS. We allow the mass, $m_{\nu}$ of the massive neutrino eigenstate in our base $\Lambda$CDM model to vary, above a lower bound of 0.06 eV. We call this model $m_{\nu}\Lambda$CDM.

Line 5 of Table \ref{tab:res} summarises the consistency tests we performed for this model.  $\sigeq$ is \seqMnu i.e. the \seqMnu 7d confidence regions only just touch, which is only a small improvement over $\Lambda$CDM. Some insight into why this happens can be gained from the bottom left panel of Fig.~\ref{fig:om_s8_multi}:
the Planck+WP contours are extended along the same line of degeneracy as the CFHTLenS constraint. 
We explain this as follows (drawing heavily on Section V of \cite{howlett12}): 
although increasing neutrino mass does reduce growth of structure, hence driving the Planck+WP contours to lower $\sigma_8$, light ($m_{\nu}$ well below 1 eV) neutrinos, are relativistic before/at recombination, so to preserve the position of the CMB acoustic peaks, $d_{A}(z_*)$ must remain constant. However, at late times, the massive neutrinos become non-relativistic, increasing the energy density relative to a model with massless neutrinos, and decreasing $d_{A}(z_*)$, unless $h$ also decreases. The degenerate combination $\Omega_mh^2$ is also well-constrained, so $\Omega_m$ must increase. 
We conclude that the datasets are still 
in tension, 
and discuss other related analyses in Section~\ref{sec:compare}.

Our principal component analysis gives a 
similar 
power law slope for this model,
and we find the constraint
\begin{equation}
\sigma_8(\Om/0.27)^{0.46} = 0.753\pm0.039,
\end{equation}
for CFHTLenS alone in the $m_{\nu}\Lambda$CDM model. This power law can be used to approximate the CFHTLenS constraints when a varying active neutrino mass is allowed.



\begin{figure}
\includegraphics[width=\linewidth]{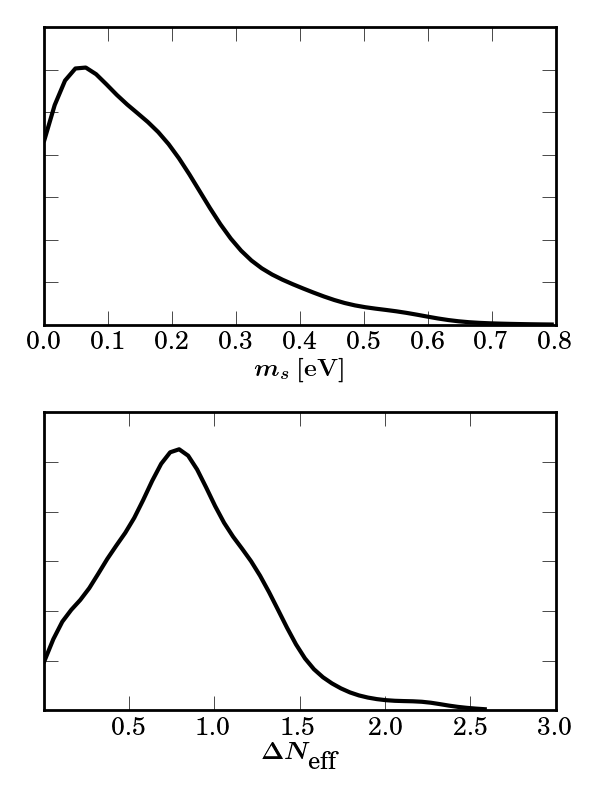}
\caption{Constraints on sterile neutrino effective mass and number of extra neutrino species from combining Planck+WP and CFHTLenS.}
\label{fig:nu_s}
\end{figure}

\begin{figure}
\includegraphics[width=\linewidth]{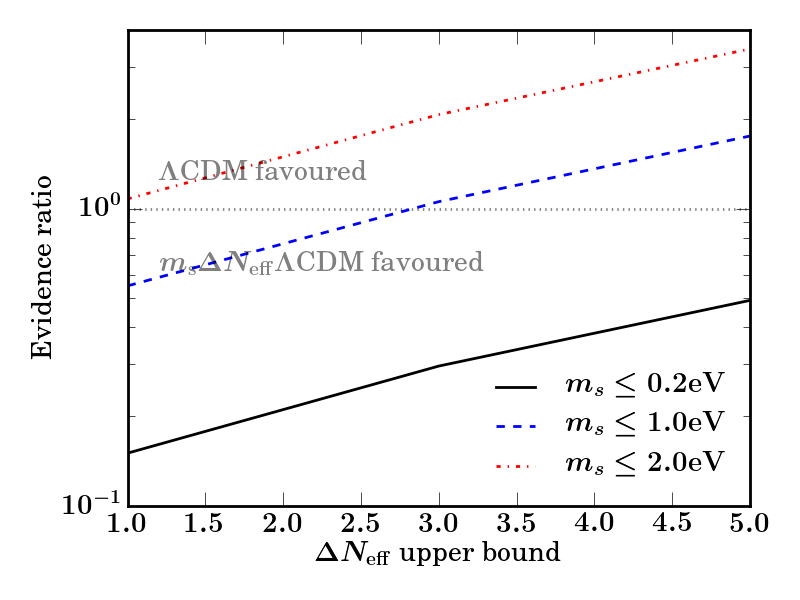}
\caption{
Evidence ratio as a function of the upper bound of the prior on $m_{\rm{s}}^{\rm{eff}}$ and $\Delta N_{\textrm{eff}}$ (both of which are assumed to have a uniform prior with a lower bound of zero). The extended model is favoured when the evidence ratio is less than one.}
\label{fig:nu_s_evidence}
\end{figure}

\subsection{A sterile neutrino: $m_{\rm{s}}^{\rm{eff}}\Delta N_{\textrm{eff}}\Lambda$CDM}\label{sec:nus}

We add to our base model a sterile neutrino - an additional neutrino species with effective mass $m_{\rm{s}}^{\rm{eff}}$ and contribution to $N_{\rm eff}$ of $\Delta N_{\rm eff} = N_{\rm eff} - 3.046$, as proposed by \cite{hamann13} and
\cite{battye14}. The bottom right panel of Fig.~\ref{fig:om_s8_multi} shows constraints in the $\Om$ - $\sigma_8$ plane, and we see now there is better agreement between the two probes - the 
68\% 2d marginalised contours are now close to touching. 
We find $\sigeq = $\seqMs for this model, a considerable improvement. We consider this agreement good enough to combine the measurements, and find $m_{\rm{s}}^{\rm{eff}}<0.408 \:\rm{eV}$(95\%) and $\Delta N_{\rm eff} = 0.819^{+0.397}_{-0.455}$. 
The 1d marginalised pdfs are shown in Fig.~\ref{fig:nu_s},
and the cosmological constraints are shown in Table \ref{tab:nus_constraints}.

\cite{leistedt14} used the Bayesian evidence ratio to assess whether extensions to $\Lambda$CDM were justified. The Bayesian evidence ratio 
for a model $M_0$ nested within model $M_1$ which has extra parameter(s) $p$ is given by 
\begin{equation}\label{eq:evrat}
\frac{{\rm P}(d | M_0 ) }{{\rm P}(d | M_1 )} = \frac{ {\rm P}(d|M_1, p=0)}{\int dp {\rm P}(d|M_1, p) {\rm Pr}(p | M_1)}
\end{equation}
where $P(d|M_1, p)$ is the likelihood of the data $d$ marginalised over all other parameters apart from $p$ and $Pr(p | M_1)$ is the normalised prior on $p$ (e.g. see \cite{lewis10}, \cite{trotta05} and references therein). 
The extended model $M_1$ is favoured if the ratio is less than one. The Jeffrey's scale \citep{jeffreys61} is often used to guide the interpretation of Baysian evidence. On this scale an evidence ratio of $\frac{1}{3}$ to $\frac{1}{10}$ would be considered substantial evidence for the extended model, while a value less than $\frac{1}{10}$ would be considered strong evidence.
If regions where the likelihood ($P(d|M_1, p)$) 
is very small are allowed by a wide prior
($Pr(p | M_1)$), 
the denominator can become very small, causing the extended model to be disfavoured. So the evidence ratio is very sensitive to the choice of prior. 

To illustrate this, we compute the evidence ratio of $m_{\rm{s}}^{\rm{eff}}\Delta N_{\textrm{eff}}\Lambda$CDM compared to $\Lambda$CDM, as a function of the priors on $m_{\rm{s}}^{\rm{eff}}$ and $\Delta N_{\rm eff}$, shown in Fig.~\ref{fig:nu_s_evidence}.
Either model can be favoured, depending on the choice of prior. If the number of extra neutrino species is assumed to be less than $2.5$ then the sterile neutrino model is favoured if we assume an upper limit on the sterile neutrino mass of $1$ eV. More stringently, if the number of extra species is assumed to be less than $2$ and the mass less than $0.2$ eV then the sterile neutrino model is around a factor of five more probable than $\Lambda$CDM. Conversely, if the prior range on the mass and number of neutrino species is large then the sterile neutrino model is disfavoured. For example, if the mass is restricted to be less than $2$ eV and the number of extra species less than 5, then $\Lambda$CDM is a little over three times as probable as $m_{\rm{s}}^{\rm{eff}}\Delta N_{\textrm{eff}}\Lambda$CDM. We note that no choice of priors considered here produces strong evidence according to the Jeffrey's scale.

Again, we provide a power law representation of the CFHTLenS constraint, finding
\begin{equation}
\sigma_8(\Om/0.27)^{0.47} = 0.750\pm0.037.
\end{equation}
This is close to the result we found for the $m_{\nu}\Lambda$CDM, which is reasonable since low redshift probes like weak lensing are sensitive to the total neutrino mass, and not to $N_{\textrm{eff}}$ (i.e. they do not care whether the neutrino mass eigenstate is active or sterile). 


\begin{table}
\centering
\begin{tabular}{ r | c | c } 
{\it Model} & Base $\Lambda$CDM & $m_{\rm{s}}^{\rm{eff}}\Delta N_{\textrm{eff}}\Lambda$CDM\\
\hline
& Planck+WP & Planck+WP \\
{\it Data} &  & CFHTLenS  \\
\hline
\hline
$\Omega_{\rm{m}}$ & $0.315^{+0.016}_{-0.018}$ & $0.274^{+0.017}_{-0.017}$ \\ 
\hline
$\sigma_8$ & $0.829^{+0.012}_{-0.012}$ & $0.811^{+0.030}_{-0.028}$ \\ 
\hline
$h_0$ & $0.673^{+0.027}_{-0.025}$ & $0.741^{+0.020}_{-0.041}$ \\ 
\hline
$n_s$ & $0.960^{+0.007}_{-0.007}$ & $0.995^{+0.014}_{-0.014}$ \\ 
\hline
$\tau$ & $0.089^{+0.012}_{-0.014}$ & $0.099^{+0.017}_{-0.017}$ \\ 
\hline
$\Delta N_{\textrm{eff}}$ & - & $0.819^{+0.397}_{-0.455}$ \\ 
\hline
$m_{\rm{s}}^{\rm{eff}}\:\rm{[eV]}$ & - & $<0.408$~(95\%)  \\ 
\hline

\end{tabular}
\caption{Cosmological parameter constraints in the $m_{\rm{s}}^{\rm{eff}}\Delta N_{\textrm{eff}}\Lambda$CDM model. The values shown are means of the posterior distribution; errors are 68\% confidence intervals unless specified. The Planck+WP base $\Lambda$CDM are included for easy reference.}
\label{tab:nus_constraints}
\end{table}

\begin{figure*}
\centerline{\includegraphics[width=20cm]{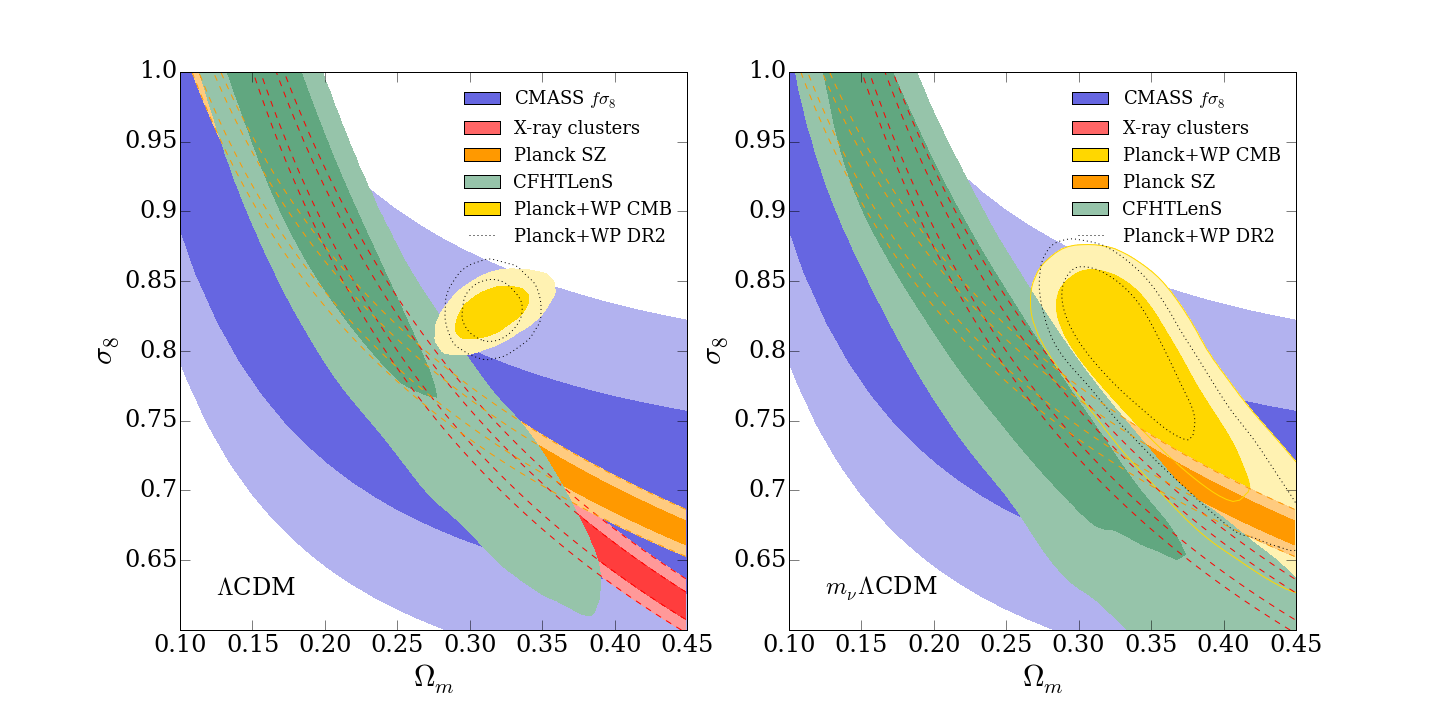}}
\caption{Comparison of constraints in the $\sigma_8$, $\Omega_{\rm m}$ plane in $\Lambda$CDM from CFHTLenS (this work; green), Planck+WP (yellow, \protect\cite{planck_cosmopar}), Planck SZ cluster counts (orange, \protect\cite{planck_sz}), X-ray clusters (red, \protect\cite{vikhlinin09}), CMASS $f\sigma_8$ (blue, \protect\cite{beutler14a}) and Planck 2015 (grey dotted, TT + lowP, \protect\cite{planckcosmo15}). In the left panel, the contours are obtained assuming $\Lambda$CDM, while in the right panel, the CFHTLenS and Planck+WP constraints allow a varying active neutrino mass. Of note is the improved consistency of the Planck+WP contours with the CMASS $f\sigma_8$ and the Planck SZ contours when the neutrino mass is allowed to vary, driving the neutrino mass detections of \protect\cite{battye14} and \protect\cite{beutler14b}.}
\label{fig:comparison}
\end{figure*}

\subsection{Primordial Gravity Waves}\label{sec:r}

Inspired by the recent BICEP2 results \citep{bicep214} we investigate the effect of gravity waves on the tension between Planck+WP and CFHTLenS. Although gravity waves are no longer required by the data \citep{planckbicep15} we still consider their implications here. Qualitatively we might expect the agreement between Planck+WP and CFHTLenS to improve due to gravity waves: the increase of power at low multipoles in the CMB from tensors \citep{crittenden93} will need to be compensated by a reduction in power from scalar modes; a reduction in scalar power would bring the Planck+WP contours closer to CFHTLenS. Meanwhile, 
the addition of tensor modes in the primordial power spectrum does not affect the matter power spectrum, which determines the shear-shear correlation function, and so has no effect on weak lensing. 
However, a detailed analysis is necessary to see how the values of the other cosmological parameters are affected by the change in shape of the CMB power spectrum due to the addition of tensors. 

The original BICEP2 measurement of $r=0.20_{-0.05}^{+0.07}$ \citep{bicep214} is not compatible with the Planck+WP data unless an additional modification is made to $\Lambda$CDM,
because 
\cite{planck_cosmopar} showed that when only $r$ is added to the base $\Lambda$CDM model, Planck+WP gives the constraint $r<0.11$ (95\% confidence).
Therefore in this section we also consider the effect of adding a running spectral index of the primordial power spectrum,
$\alpha_{run}=\text{d}n_s/\mathrm{d}lnk$, which \cite{planck_cosmopar} showed to relax the constraint on $r$ to $r<0.26$ (95\%),
into agreement with BICEP2. 

We repeat the investigation in $n$-dimensions described in \ref{sec:tension}, and find that the addition of gravity waves and running of the spectral index, in the $r\alpha_{run}\Lambda$CDM model, relaxes the tension between the datasets only slightly, with $\sigma_{\textrm{eq}}=$\seqRNrun i.e. the \seqRNrun iso-likes from each dataset touch.

Therefore 
we conclude that gravity waves do not significantly resolve the tension between CFHTLenS and Planck+WP.





\section{Discussion}

In this Section we compare our results with those from other analyses, and speculate on alternative potential explanations for the discrepancy.
We will refer to Fig.~\ref{fig:comparison}, which shows
a selection of other low-z probes of 
the growth of structure. 

\subsection{Comparison with other work}\label{sec:compare}

Several other authors have considered how to reconcile cosmology from the CMB and the amplitude of matter fluctuations measured by low-redshift probes. The most relevant to our work are by 
\cite{planck_cosmopar},
\cite{battye14},
\cite{beutler14b},
\cite{dvorkin14},
\cite{leistedt14},
and
\cite{archidiacono14}.
We discuss next the differences to our analysis.

The \cite{planck_cosmopar} noted an approximately $2\sigma$ discrepancy between their Planck CMB analysis and the CFHTLenS analysis of 
\cite{heymans13} and noted that further work will be required to resolve the difference. They allow freedom in the effect of lensing on the primary anisotropies and find that a larger lensing amplitude is preferred when the Planck data is combined with smaller scale CMB measurements.
Taken at face-value this suggests an increased $\sigma_8$ from low redshift data, unlike all the other low redshift data considered in the other papers we discuss below.

The tension between Planck Sunyaev--Zel'dovich (SZ) cluster counts and the primary anisotropies was discussed by the \cite{planck_sz}. They discuss possible systematics in the SZ analysis and conclude that each is improbable, but that understanding the mass bias scaling relation is the key to further investigation. They find a $1.9\sigma$ preference for a non-zero active neutrino mass by combining Planck+WP with the Planck SZ constraints, marginalising over their preferred range in the hydrostatic mass bias ($0.7<1-b<1.0$). 
Two recent analyses have attempted to constrain the mass bias by comparing the Planck mass estimates ($M_{Planck}$) with weak-lensing mass estimates ($M_{WL}$). Using the ratio $<M_{Planck}/M_{WL}>$ as a proxy for the mass bias, \cite{vonderLinden14} find $<M_{Planck}/M_{WL}>=0.688\pm0.072$ for 22 of the clusters in the Planck cosmology sample, and note that adopting this mass bias would substantially reduce the tension. \cite{hoekstra15} find $<M_{Planck}/M_{WL}>=0.76\pm0.05 \rm{(stat)} \pm 0.06 \rm{(sys)}$ for 37 clusters common between the Canadian Cluster Comparison Project and the Planck sample, and conclude that this does not resolve the tension.

\cite{planck_lensing} used lensing of the CMB to measure the power spectrum of the gravitational potential at slightly higher redshift than that probed by CFHTLenS. This was combined with the constraints from the primary anisotropies and found to reduce the measured amplitude of fluctuations \citep{planck_cosmopar}. 
One of the many extensions to $\Lambda$CDM investigated by the Planck team~\citep{planck_cosmopar} was the mass of the active neutrino. 
When using the CMB lensing information, they found that this increased the upper limit on the neutrino mass relative to that from CMB primary anisotropies alone (the 95\% upper limit increased from 0.66 eV to 0.85 eV), indicating some tension. 

\cite{battye14} found a preference for a non-zero active neutrino mass when combining CMB lensing, CFHTLenS and Planck SZ cluster counts \citep{plancksz13} with the CMB. 
They use the correlation functions measured by \cite{kilbinger13}, who performed a 2d cosmic shear analysis i.e. they did not use multiple redshift bins. 
They found similar but stronger preference for a non-zero sterile neutrino mass. They noted that both these joint fits come at the cost of a worse fit to the primary CMB data. 
Fig.~\ref{fig:comparison} shows an orange band corresponding to the SZ cluster counts prior from \cite{plancksz13}, $\sigma_8(\Om/0.27)^{0.3} = 0.78\pm0.01 (68\%)$. The shallower degeneracy direction of this prior as compared to the lensing constraint allows more overlap with the Planck+WP confidence regions with an active neutrino, explaining the significant detection of neutrino mass \cite{battye14} claimed when combining this prior with Planck+WP.

\cite{hill14} also used SZ information from Planck, constructing a thermal SZ map and cross-correlating with the Planck CMB lensing potential map. They constrain $\sigma_8(\Omega_{\rm{m}}/0.282)^{0.26} = 0.824 \pm 0.029$, a result consistent (within $\Lambda$CDM) with that from the Planck primary aniotropies, unlike the aforementioned SZ cluster counts.

\cite{beutler14b} investigate the constraints on the active neutrino mass using the Baryon Oscillation Spectroscopic Survey (BOSS, \cite{schlegel09}), the CMB, and other low redshift measurements including a parameterised fit to the CFHTLenS cosmology constraints of \cite{kilbinger13}.
Their Figure 7 illustrates that the inclusion of a free active neutrino mass elongates the Planck contours in a direction parallel to the CFHTLenS constraints.
They combine CMB constraints with those from the BOSS
CMASS DR11 galaxy clustering results of \cite{beutler14a}, which come from BAO, Alcock-Paczinski and growth measurements. The push towards a positive neutrino mass comes mostly from the growth constraint (shown as blue contours in Fig.~\ref{fig:comparison}) since this is sensitive to the amplitude of clustering. They use various combinations of the data and find similar non-zero values for the neutrino mass to \cite{battye14}, and finally combine them all together to get a $\approx3\sigma$ detection of the neutrino mass, with a similar result whether using WMAP or Planck temperature anisotropies.

\cite{dvorkin14} focus on the discrepancy between Planck and BICEP2, noting that extra relativistic species in the early universe can help alleviate the tension introduced into the Planck data by extra power from gravitational waves. They point out that the sterile neutrino can thus help alleviate the Planck-BICEP2 tension and additionally the CMB-low-z tension at the same time. They use local $H_0$, baryon acoustic oscillations and local X-ray cluster abundance measurements (\cite{vikhlinin09}, shown as the red band in Fig.~\ref{fig:comparison}) for low-redshift information, and obtain a $\approx3\sigma$ sigma detection of the sterile neutrino mass. 

The cosmological constraints on sterile neutrinos are compared with those from short baseline neutrino oscillation experiments in \cite{archidiacono14}. They use the CMB combined with low-redshift clustering measurements from the growth of structure obtained by \cite{parkinson12}, Planck SZ and the parameterised fit to the CFHTLenS constraints of \cite{kilbinger13}. 
They find a detection of non-zero sterile neutrino mass, and
point out the significant inconsistency between its value and that found in the neutrino oscillation experiments. 

However, the neutrino mass detections are disputed by \cite{leistedt14}, who point out that they are driven by two highly constraining datasets from counting galaxy clusters \citep{plancksz13, vikhlinin09}. They describe some of the potential systematic effects in these two measurements. They use the same simple parameterised fit to the CFHTLenS constraints as \cite{archidiacono14} and \cite{beutler14b}. Omitting the datasets in tension, they use each other low redshift probe one-at-a-time, and find only upper limits on the neutrino mass.

Other recent cosmic shear analyses have shown some variation in the preferred value of $\sigma_8$, which likely indicates their consistency with Planck. \cite{kitching14} performed a `3D cosmic shear' analysis of CFHTLenS with conservative cuts on the scales considered ($k \le 1\: \rm{hMpc^{-1}}$), leading to larger statistical errors, but probably less systemtic error due to uncertainty in the nonlinear matter power spectrum, and found constraints consistent with Planck+WP. \cite{jee13} use $\xi_{+/-}$ measurements from the Deep Lens Survey, and found $\sigma_8$ at $\Omega_m=0.3$ to be $0.804\pm0.21$, significantly higher than the value of $0.73\pm0.035$ from this work. However, we note that their use of angular scales down to $0.3"$ in both $\xi_+$ and $\xi_-$, and the \citetalias{smith03} version of {\sc Halofit} (which predicts less power at small scales than the version used in this work) are likely to contribute to this.

The Planck 2015 results (in particular \cite{planckcosmo15}, Planck 15 henceforth) released since submission of this work, appear to have changed little of the above conclusions (as suggested by the grey dotted contours in Fig.~\ref{fig:comparison}). The tension in $\sigma_8$ with a larger catalog of SZ clusters \citep{plancksz15} remains, as does the uncertainty on the mass calibration. The preference for higher lensing smoothing in the CMB temperature power spectrum (suggesting a larger amplitude of fluctuations) also remains, as does the the lower amplitude preferred by the lensing reconstruction data \citep{plancklensing15}. The combined constraint on the neutrino mass from the primary temperature anisotropies, LFI polarisation and CMB lensing is now $\Sigma m_{\nu} < 0.68\; \rm{eV}$. 


\subsection{Other possible explanations}

We have shown our results to be robust to uncertainties in the modelling of the nonlinear matter power spectrum (including AGN feedback) and intrinsic alignments. Two other weak-lensing systematics we have not considered in detail are photometric redshift errors and shape measurement errors. We can estimate how wrong these would have to be to account for the $\approx20\%$ disparity (assuming the Planck+WP best-fit value of $\Omega_m$) between the CFHTLenS and Planck+WP best fit values of $\sigma_8$. As a rule of thumb, equation 24 from \cite{huterer06} tells us that the lensing power spectrum (at $l=1000$ and assuming all source galaxies are at $z_s=1$) has dependence
\begin{equation}\label{eqn:pz_err}
P^{\kappa}\propto\sigma_8^{2.9} z_s^{1.6}. 
\end{equation}
Hence to observe the same signal (i.e. setting $P^{\kappa}$ equal to a constant in \ref{eqn:pz_err}), with a 20\% higher $\sigma_8$, would require the redshifts to shift systematically by a fraction $\approx1.2^{-1.8}=0.72$ i.e. a systematic 30\% error in photometric redshift, which would be very surprising in any one redshift bin, let alone all redshift bins with the same sign.

To estimate the effect of multiplicative bias on our results, we note that $\xi_{+/-}\propto(1+m)^2$, where $m$ is the multiplicative bias (see e.g. \cite{heymans06} for an introduction to shape measurement biases). The information in $\xi_{+/-}$ comes from a mixture of linear and nonlinear scales. On linear scales we have $\xi_{+/-}^{observed}\propto (1+m)^2 P_{\delta} \propto (1+m)^2\sigma_8^2$, whereas on nonlinear scales $\xi_{+/-} \propto (1+m)^2\sigma_8^3$. So an increase in $\sigma_8$ of 20\% would require $(1+m)=1.2^{-1}=0.83$ (assuming all information comes from linear scales) and $(1+m)=1.2^{-3/2}=0.69$ (assuming all information comes from nonlinear scales). The multiplicative bias in the CFHTLenS shape measurements was calibrated using image simulations \citep{miller13} and the average value of $(1+m)$ was found to be 0.94. It's clear then, that the value of the multiplicative bias estimated from simulations would have to be catastrophically wrong to produce such a significant shift in $\sigma_8$.

\cite{spergel13} reanalysed the Planck data, 
and claim that the $217\:\rm{GHz} \times 217\:\rm{GHz}$ detector set spectrum used in the Planck analysis is responsible for `some' of the tension with other cosmological measurements.
The latest version of \cite{planck_cosmopar} does discuss a residual systematic in the $217\:\rm{GHz} \times 217\:\rm{GHz}$ spectrum, but claim that this has an impact of less than half a standard deviation on cosmological parameters. The Planck 2014 releases will include a correction for this, but the Planck products used in this analysis do not.
Furthermore, the CMB constraints we use rely on the WMAP polarisation data primarily to constrain the optical depth to reionisation. The Planck satellite has measured the polarisation signal more precisely. A reduction in the optical depth to reionisation would be required to push the CMB contours towards those from CFHTLenS. This is because a lower optical depth increases the predicted CMB temperature anisotropy power spectrum on the majority of scales, and thus the underlying amplitude of scalar fluctuations must be reduced to retain a good fit to the observations.
The \cite{planck_cosmopar} noted that the Planck temperature anisotropies alone tightly constrain the combination
\begin{equation}
\sigma_8 e^{-\tau} = 0.753 \pm 0.011
\end{equation}
which would require $\tau$ to be significantly smaller to 
make a significant impact on the discrepancy. 
Planck 15 results did indeed find a lower value of $\tau$ of $0.078\pm0.019$ (down from $0.089\pm0.013$), however, this was accompanied by an upward shift in the overall calibration of the temperature data, which means the the value of $\sigma_8$ preferred by Planck has not changed.

As noted by \cite{archidiacono14}, the cosmology constraints on the sterile neutrino are not compatible with those from short baseline neutrino experiments. However, the cosmology constraints are relatively generic for other relativistic particles in the early universe. 

Finally we note that other extensions of $\Lambda$CDM can be 
explored, such as the variation of the dark energy equation of 
state, a universe with non-zero curvature, or a deviation of the growth of structure from the GR prediction. 

\section{Conclusions}
We have confirmed the tension between Planck+WP and CFHTLenS in the $\Omega_m$-$\sigma_8$ plane within the base $\Lambda$CDM cosmology, and shown it's robustness to various weak lensing systematics. We find that considering the overlap in the full $\Lambda$CDM parameter space weakens this conclusion, since marginalising over some of the parameters makes contours tighter (the 68\% contours lie at higher probability, and closer to the best fit point) in the remaining parameters.



We find 
that allowing massive active neutrinos
does not significantly resolve the tension, because the slope of the CFHTLenS contours runs parallel to the effect of adding active neutrinos to the CMB. Other works include other datasets with a shallower slope than CFHTLenS, which intersect the CMB contours at high active neutrino mass, thus leading to a detection of the neutrino mass. 
However, if taken at face value, the CFHTLenS and Planck+WP data already rule out this scenario. 
It was noted in \cite{battye14} that the active neutrino mass detection comes at the cost of a decreased likelihood of the Planck+WP data. In this paper we have quantified the size of this decrease in terms of the full n-dimensional contours and used a more robust version of the cosmic shear data. 

The addition of 
tensor modes, 
even with 
running of the spectral index also does not significantly affect the tension. 

We have also added 
an extra, sterile species of neutrino,
and find that the \seqMs confidence iso-likes in the 8 dimensional parameter space touch in this case.
We find that 
the effective number of extra neutrino species ($\Delta N_{\textrm{eff}}$)
is favoured to be non-zero in the joint fit, at about the $2\sigma$ level. 
Although the $m_{\rm{s}}^{\rm{eff}}\Delta N_{\textrm{eff}}\Lambda$CDM model does allow an acceptable joint fit, some tension remains between Planck+WP and CFHTLenS, 
since 
all points in the joint fit 
are 
on at least the \seqMs 8d iso-like of either the Planck+WP-only or CFHTLenS-only constraints (and at least the 68\% 2d marginalised contour in the $\Omega_m - \sigma_8$ plane). 

Therefore we are not completely satisfied by the amount which the
flexible sterile neutrino model reduces the discrepancy, 
and believe that investigating other new physics, and other sources of systematic error in either experiment, may lead to a better resolution of the tension. 

If the $\Lambda$CDM discrepancy between CFHTLenS and Planck+WP is to be resolved it would require more than a $1\sigma$ shift in the CFHTLenS constraints (on e.g. $\sigma_8(\Om/0.27)^{0.47}$) upwards or similarly for the Planck+WP downwards. We look forward to future constraints from the CMB and other lensing analyses.

\section*{Acknowledgements}

Many thanks to Catherine Heymans, Richard Battye, Boris Leistedt, Fergus Simpson, Adam Amara, Anthony Lewis, Adam Moss, Tomasz Kacprzak, George Efstathiou, Ben Wandelt, John Beacom, Ofer Lahav,  M. James Jee, J. Anthony Tyson and Michael D. Schneider for useful discussions.
Thanks also to the non-author members of the core CosmoSIS team: Marc Paterno, Elise Jennings, Douglas Rudd, Alessandro Manzotti, Scott Dodelson, Saba Sehrish, James Kowalkowski.

NM acknowledges support from the Science \& Technologies Facilities Council.
JZ and SB acknowledge support from the European Research Council in the form of a Starting Grant with number 240672. BJ is partially supported by the US Department of Energy grant de-sc0007901.
This work was supported in part by National Science Foundation Grant No. PHYS-1066293 and the hospitality of the Aspen Center for Physics

This work is partly based on observations obtained with MegaPrime/MegaCam, a joint project of CFHT and CEA/IRFU, at the Canada-France-Hawaii Telescope (CFHT) which is operated by the National Research Council (NRC) of Canada, the Institut National des Sciences de l'Univers of the Centre National de la Recherche Scientifique (CNRS) of France, and the University of Hawaii. This research used the facilities of the Canadian Astronomy Data Centre operated by the National Research Council of Canada with the support of the Canadian Space Agency. CFHTLenS data processing was made possible thanks to significant computing support from the NSERC Research Tools and Instruments grant program.

This work also uses observations obtained with Planck (http://www.esa.int/Planck), an ESA science mission with instruments and contributions directly funded by ESA Member States, NASA, and Canada.

\bibliography{refs}

\appendix
\section{Confidence levels and the number of degrees of freedom}\label{appendix:nd_contour_stuff}

To parameterise the probability distribution of one or more model parameters, the 68\% (95\%) confidence intervals, defined as the contour (in the 2d case) of constant probability that encloses 68\% (95\%) of the probability distribution, is often used. In this paper we've used these contours (or `iso-likelihood surfaces' in $>2$ dimensions) to judge the consistency of the parameter constraints from two datasets. We've seen that when marginalising over several parameters, and reducing the 6+ dimensional parameter space to 2d, the percentile values of the just-overlapping surfaces/contours increase, giving the impression of greater discrepancy. 

\begin{figure}
\includegraphics[width=\linewidth]{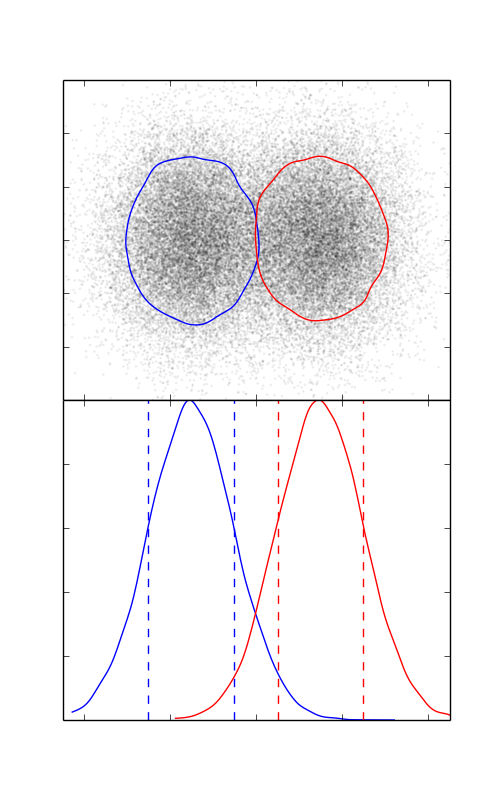}
\caption{Top panel: samples and 68\% confidence regions for 2 2d gaussian pdfs. Bottom panel: The pdfs marginalised in the y-direction, with 68\% confidence levels as vertical dashed lines.}
\label{fig:toy_pdfs}
\end{figure}

This phenomenon is illustrated by figure \ref{fig:toy_pdfs}, which shows samples from two 2d gaussian pdfs whose 68\% confidence regions touch in 2d (upper panel), but do not when the parameter in the y direction is marginalised over (lower panel). This is related to the effect described in the table on page 815 Numerical Recipes in C \citep{press07} (page 693 of the 2nd edition), which shows how the $\Delta\chi^2$ (the change in probability relative to the maximum probability point in parameter space) of the 68\% confidence level varies with $\nu$, the number of degrees of freedom (analogous to the number of dimensions). They show that the $\Delta\chi^2$ of a given confidence level increases with $\nu$, or equivalently, for a given $\Delta\chi^2$, the confidence level increases as $\nu$ is reduced. This is consistent with figure \ref{fig:toy_pdfs} - $\nu$ is reduced from 2 to 1 by marginalising, the $\Delta\chi^2$ of the 68\% level increases, so is found closer to the peak, and the constraint appears tighter. This tightening when reducing $\nu$ is consistent with the higher apparent significance of the tension in the 2d marginalised plots throughout this work, as compared to the quoted n-dimensional $\sigma_{eq}$ values.

\end{document}